\documentclass{article}
\usepackage{hiph-art}
\usepackage{epsf}
\usepackage{graphicx}
\volnumber{19} \issuenumber{1} \edyear{2005}                             
\frompage{000} \topage{000}                                              
\recrevdate{30 October 2005}                                              
\def\rightmark{Thermal Model}
\def\leftmark{J. Cleymans, H. Oeschler, K. Redlich, S. Wheaton}
\title{The Thermal Model and the Transition from Baryonic to Mesonic Freeze-Out.}
\authors{
{J. Cleymans$^1$, H. Oeschler$^2$, K, Redlich$^3$, S. Wheaton$^1$ %
\index{Cleymans,J.} 
\index{Oeschler, H.} 
\index{Redlich, K.} 
\index{Wheaton, S.} 
}\\[2.812mm]
{\normalsize
\hspace*{-8pt}$^1$ UCT-CERN Research Centre and Department  of  Physics,
University of Cape Town, Rondebosch 7701, South Africa,\\[0.2ex] 
\hspace*{-8pt}$^2$ Darmstadt University of Technology, D-64289 Darmstadt, Germany\\[0.2ex]
\hspace*{-8pt}$^3$Institute of Theoretical Physics, University of Wroc{\l}aw,\\
 Pl. Maksa Borna 9, 50-204  Wroc{\l}aw, Poland,\\
and CERN TH, CH 1211 Geneva 23, Switzerland
}}
 
\abstract{ The present status of the thermal model is reviewed and
the recently discovered sharp peak in the $K^+/\pi^+$ ratio is discussed
in this framework. 
It is shown that the rapid change is related to a 
transition from a baryon dominated hadronic gas to a 
meson dominated one.  Further experimental tests to clarify the nature of the 
transition are discussed. In the thermal model the  corresponding maxima 
in the $\Xi/\pi$ and $\Omega/\pi$ ratios occur at slightly different beam 
energies.}

 
\makeindex
\begin{document}
 
\newcommand{\eovern}{\left<E\right>/\left<N\right>}
\maketitle

\section{Introduction}
It is by now well-known that particle yields integrated over all momenta
can be calculated as if they originate from a fireball at rest, provided
chemical freeze-out happens at the same temperature and chemical 
potential on all points of the freeze-out surface.
This follows from a covariance argument using the Cooper-Frye~\cite{cooper-frye} formula as a
starting point.
The number of particles of type $i$ is determined by:
$$
E{dN_i\over d^3p}  = {g_i \over (2\pi)^3}\int  d\sigma_\mu p^\mu  \exp\left({-{p^\mu u_\mu\over T} + {\mu_i\over T}}\right)
$$
where $u^\mu$ is the flow velocity which, in general, depends on space and time.
Integrating this over all momenta leads to,
$$
N_i  = {g_i \over (2\pi)^3}\int d\sigma_\mu 
\int {d^3p\over E}p^\mu \exp\left({-{p^\mu u_\mu\over T} + {\mu_i\over T}}\right) .
$$
Lorenz invariance dictates that the integral over momenta depends only on the flow four-velocity $u^\mu$. Hence
$$
N_i  = \int d\sigma_\mu u^\mu n_i(T,\mu )
$$
where $n_i(T,\mu)$ is the particle density in a fireball at rest.
 If the temperature and chemical potential are unique along the freeze-out curve
then $n_i$ has the same value along the freeze-out curve and one obtains,
$$
N_i  = n_i(T,\mu )\int d\sigma_\mu u^\mu  ,
$$
i.e. integrated ($4\pi$)  multiplicities are the same as for a single fireball at rest. 
The volume has, of course, a completely different 
interpretation and depends in a complex manner
on the evolution of the fireball.

\section{Present Knowledge of the Chemical Freeze-Out Diagram}
\begin{figure}[th]
\insertplot{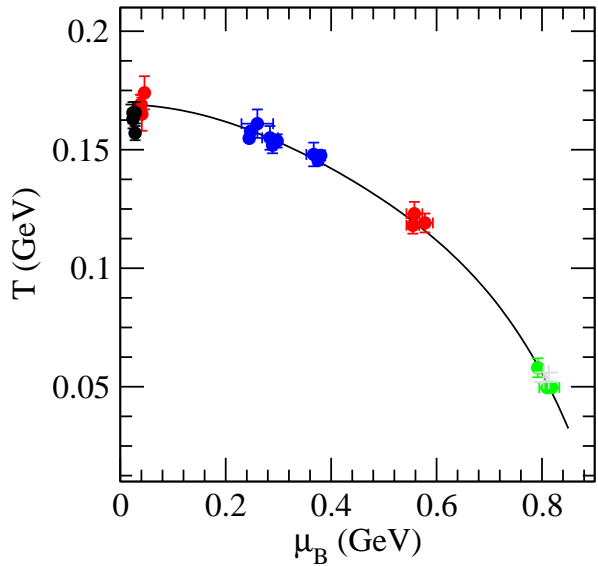}
\caption[]{Values of $\mu_B$ and $T$ for different energies obtained from RHIC, SPS, AGS 
and SIS. The solid line  has been drawn to guide the eye.}
\label{fig1}
\end{figure}

The present knowledge of the freeze-out parameters in relativistic heavy ion collisions
is summarized in Fig.(\ref{fig1}) and shows a remarkable regularity when going from
the lowest to the highest energies~\cite{cr1,cr2}. 
\section{The Horn in the $K^+/\pi^+$ Ratio}

The NA49 Collaboration  has recently  performed a series of measurements
of Pb-Pb collisions at 20, 30, 40, 80 and 158 AGeV beam energies~\cite{gazdzicki}.
When these results are combined with measurements at lower beam energies
from the
AGS they
reveal an unusually sharp variation with beam energy 
in the $\Lambda/\left<\pi\right>$, 
with
$\left<\pi\right>\equiv 3/2(\pi^++\pi^-)$,
 and $K^+/\pi^+$ ratios. Such a strong variation with
energy does not occur in pp collisions and therefore indicates a
major difference in heavy-ion collisions. This transition
 has been referred  as the ``horn''.

To understand this behavior, which is not seen anywhere in $p-p$ or $e^+-e^-$
collisions, 
one uses 
 the Wroblewski factor~\cite{wroblewski},
$$
\lambda_s = {2\left<s\bar{s}\right>\over \left<u\bar{u}\right> + \left<d\bar{d}\right>} ,
$$
which compares 
the number of \underline{{\bf newly}} created 
quark-anti-quark pairs
\underline{{\bf before}} strong decays, i.e. before $\rho$'s and $\Delta$'s decay. The 
limiting values are:
$\lambda_s = 1$ all quark pairs are equally abundant, 
and $\lambda_s = 0$ if  no strange quark pairs are present.\\

\begin{figure}[htb]
\insertplot{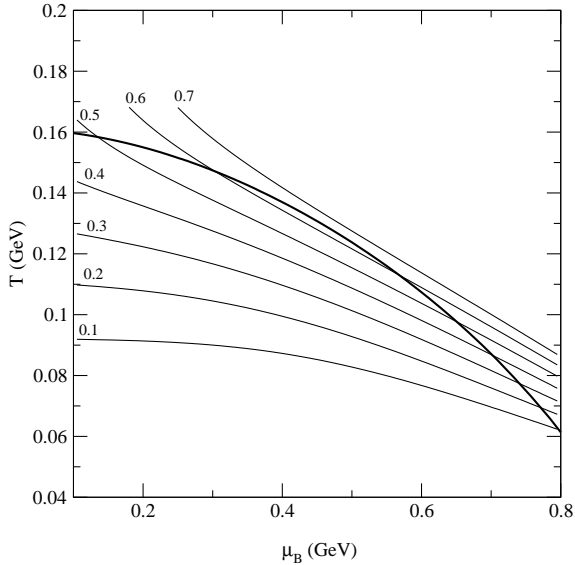}
\caption[]{Lines corresponding to fixed values of the Wroblewski ratio.
The thick solid line is the freeze-out curve corresponding to 
$\eovern$ =  1 GeV.}
\label{fig2}
\end{figure}

Its behavior in the thermal model is shown in Fig.(\ref{fig2}) where lines 
corresponding to fixed values of the Wroblewski factor are shown. Small values
correspond to low values of the temperature and the dependence on the chemical
potential is not very pronounced. The solid line in Fig.(\ref{fig2}) corresponds to
a freeze-out curve with $\eovern$ = 1 GeV. At low energies, following
the freeze-out curve, a smooth increase in the Wroblewski factor arises but
around AGS energies, the Wroblewski factor reaches a maximum which is followed by
a steady decrease towards RHIC energies. The relative strangeness content at AGS energies 
is {\bf larger} than at RHIC energies.  This is the basis for the explanation of the 
``horn'' in the thermal model.
\section{Explanation for the Horn}

\begin{figure}[htb]
\insertplot{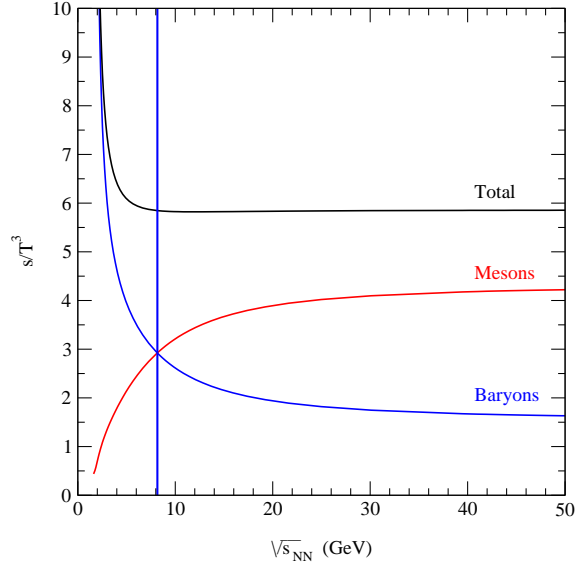}
\caption[]{
The entropy density normalised to $T^3$
as a function of the beam energy as calculated in the thermal model.
The contributions from baryons and mesons are shown separately.}
\label{fig3}
\end{figure}

The change in the Wroblewski factor is accompanied by a change in the composition of 
the hadronic fireball. Below the maximum in the Wroblewski ratio, the fireball
 is dominated by baryons, above it is dominated by mesons. To determine the precise position of
the change-over we use the entropy density, normalized to $T^3$. This is shown 
in Fig.(\ref{fig3}). 

In the statistical model~\cite{horn} therefore
a rapid change is expected as the hadronic gas undergoes a
transition from a baryon-dominated  to a meson-dominated gas. The
transition occurs at a temperature $T$ = 140 MeV and baryon
chemical potential $\mu_B$ = 410 MeV corresponding to an incident
energy of $\sqrt{s_{NN}}$ = 8.2 GeV. 

\begin{figure}[htb]
\insertplot{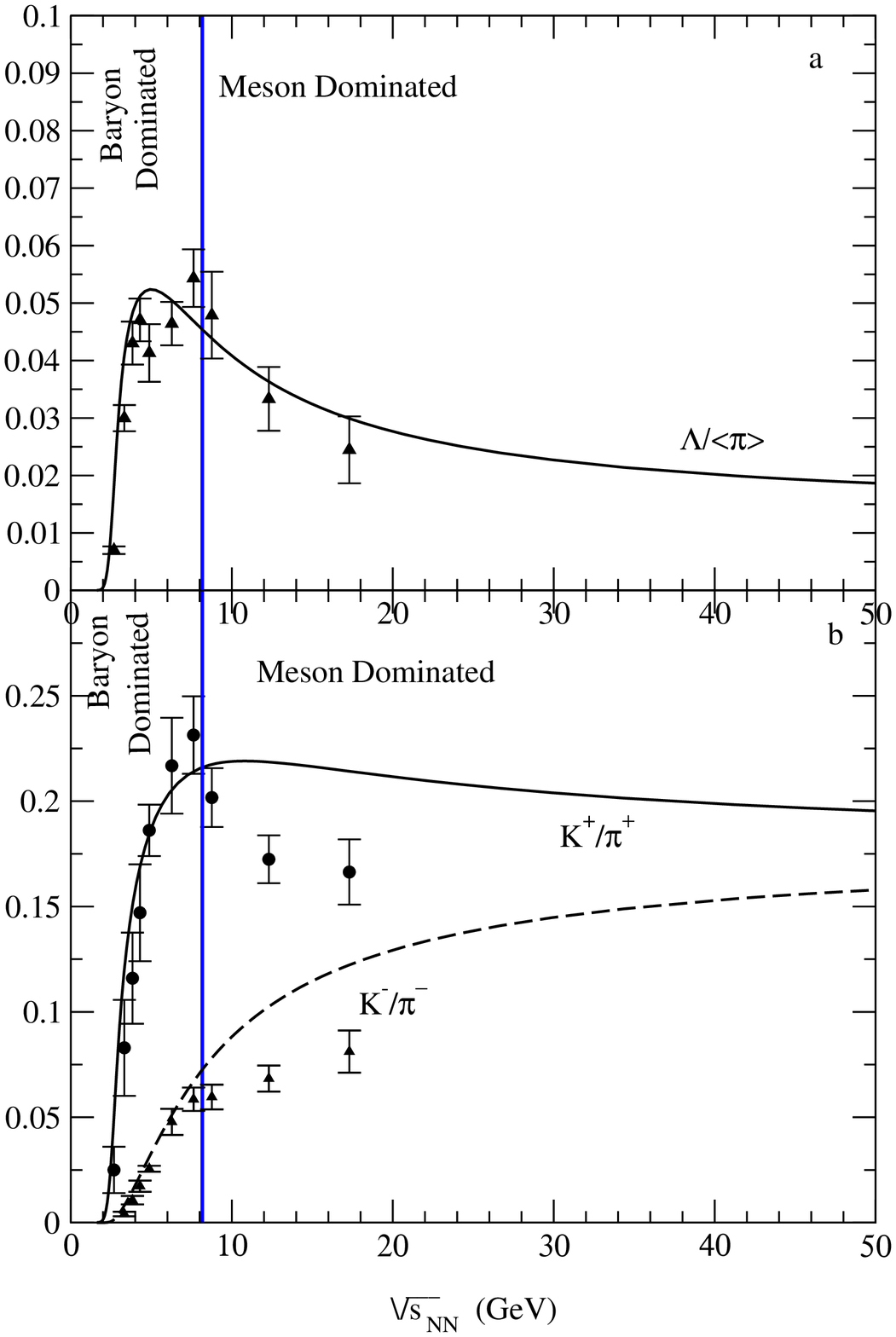}
\caption[]{(a) The $\Lambda/\left<\pi\right>$ ratio as a function of energy.
(b) The $K^+/\pi^+$ and $K^-/\pi^-$ ratios as a function of energy. The solid and dashed lines are the results of the thermal model.}
\label{fig4}
\end{figure}

It is to be expected that if the maxima observed in the particle ratios
do not all occur at the
same temperature, i.e. at the same beam energy, then the case for
a phase transition is not very strong.
A comparison with the data is shown in Fig.(\ref{fig4}).

The observed behavior seems to be governed by properties of the hadron
gas. 
In order to distinguish between this smooth change-over from a baryon-dominated
to a meson-dominated hadronic gas it would be useful to determine the 
positions of the maxima more precisely. In the thermal model these are 
given in Table I.

More detailed experimental studies of multi-strange hadrons
will allow the verification or disproval of the trends shown in this
paper.
It should be clear that the 
$\Omega^-/\pi^+$ ratio is very broad and shallow and it  will be difficult 
to find a maximum experimentally.

\begin{table}
\begin{center}
\begin{tabular}{|l|c|c|}
\hline
Ratio                       &Maximum at            & Maximum   \\
                            &$\sqrt{s_{NN}}$ (GeV)  & Value  \\
\hline                                                         
                            &                       &           \\
$\Lambda/\left<\pi\right>$  & 5.1                   & 0.052     \\
$\Xi^-/\pi^+$               & 10.2                  & 0.011     \\
$K^+/\pi^+$                 & 10.8                  & 0.22      \\
$\Omega^-/\pi^+$            & 27                  & 0.0012     \\
\hline
\end{tabular}
\caption{\label{table} Maxima in particle ratios as
 predicted by the thermal model. }
\end{center}
\end{table}


\section{Summary}
 In conclusion, while  the statistical model cannot explain
the sharpness of the peak in the $K^+/\pi^+$ ratio, its position
corresponds  precisely to a transition from a baryon-dominated to
a meson-dominated hadronic gas. This transition occurs at a
\begin{itemize}
\item temperature $T = $ 140 MeV, 
\item baryon chemical potential $\mu_B = $ 410 MeV, 
\item energy $\sqrt{s_{NN}} = $ 8.2 GeV. 
\end{itemize}
In the
statistical model this transition leads to a sharp peak in the
$\Lambda/\left<\pi\right>$ ratio, and to
 moderate peaks in the $K^+/\pi^+$, $\Xi^-/\pi^+$ and
$\Omega^-/\pi^+$ ratios. Furthermore, these peaks are at 
different energies in the statistical model. The statistical model
predicts that the maxima in the $\Lambda/\left<\pi\right>$,
$\Xi^-/\pi^+$ and $\Omega^-/\pi^+$ occur at increasing
beam energies. 

\section*{Acknowledgments}
Two of us (J.C. and S.W.) would like to thank the theory division
of the GSI for their hospitality.  The  partial  support by the
Polish Committee for Scientific Research under contract 2P03
(06925) and the Polish--South--African research project is
acknowledged. On of us (J.C.) acknowledges financial support of the
Alexander von Humboldt foundation.

\vfill\eject

\end{document}